\documentclass[aps,preprint,superscriptaddress,letterpaper]{revtex4}
\usepackage{graphicx,amssymb,bm,natbib,amsfonts,amsmath}

\usepackage{hyperref}

\begin{document}

\title{Three-Fold Diffraction Symmetry in Epitaxial Graphene and the SiC Substrate}

\author{D.A. Siegel}
\affiliation{Department of Physics, University of California,
Berkeley, CA 94720, USA}
\affiliation{Materials Sciences Division,
Lawrence Berkeley National Laboratory, Berkeley, CA 94720, USA}

\author{S.Y. Zhou}
\affiliation{Department of Physics, University of California,
Berkeley, CA 94720, USA}
\affiliation{Materials Sciences Division,
Lawrence Berkeley National Laboratory, Berkeley, CA 94720, USA}

\author{F. El Gabaly}
\affiliation{Sandia National Laboratories, Livermore, California 94551, USA}

\author{A. K. Schmid}
\affiliation{Materials Sciences Division,
Lawrence Berkeley National Laboratory, Berkeley, CA 94720, USA}

\author{K. F. McCarty}
\affiliation{Sandia National Laboratories, Livermore, California 94551, USA}

\author{A. Lanzara}
\affiliation{Department of Physics, University of California,
Berkeley, CA 94720, USA}
\affiliation{Materials Sciences Division,
Lawrence Berkeley National Laboratory, Berkeley, CA 94720, USA}

\date{\today}

\begin{abstract}

The crystallographic symmetries and spatial distribution of stacking domains in graphene films on SiC have been studied by low energy electron diffraction (LEED) and dark field imaging in a low energy electron microscope (LEEM).  We find that the graphene diffraction spots from 2 and 3 atomic layers of graphene have 3-fold symmetry consistent with AB (Bernal) stacking of the layers.  On the contrary, graphene diffraction spots from the buffer layer and monolayer graphene have apparent 6-fold symmetry, although the 3-fold nature of the satellite spots indicates a more complex periodicity in the graphene sheets.

\end{abstract}

\maketitle


The past few years have witnessed a growing need to identify new methods for the synthesis of graphene films for both basic research and industrial applications.  Of all the methods explored so far, substrate growth methods seem the most promising due to the ease and reliability of growth of large-scale films \cite{oldestnickel,Sutter,Berger}.
But the choice of substrate often has a major impact on the properties of the graphene film.
Graphene on hexagonal boron nitride, AB-stacked bilayer graphene, and graphene grown on the C face of SiC (which possesses azimuthal rotations between layers), are typical examples of how the stacking of the graphene film on top of a substrate can lead to the breaking of the six fold symmetry inducing the opening of a gap in the electronic structure \cite{KAWASAKI,Giovannetti,OhtaBilayer, Blase, ZhouGap, Kim}, or alternatively to a full decoupling between different layers \cite{Berger,HassReview,Varchonsimulation,CONRADPAPER}.

Therefore, understanding how graphene grows on top of a substrate is fundamental to engineering new graphene sheets with controlled properties.
Here we will focus on epitaxial graphene grown on the Si face of SiC, one of the most studied graphene systems because of its potential for industrial application due to the presence of a bandgap in measured as well as calculated spectra \cite{ZhouGap, Kim}.  The mechanism behind this gap opening is still under debate, so understanding the precise structure of the graphene/buffer layer system remains an important issue \cite{Brihuega, Varchon}.

One way to answer questions about the structure of this material might be through low energy electron diffraction (LEED), which is a more direct probe of the crystal symmetry than STM \cite{Tománek,Brihuega,Varchon} provided that the diffraction can be performed with a spatial resolution that is smaller than the structural domains of the crystal.  For example, LEED from a single Ru(0001) terrace has the 3-fold symmetry of the hcp layer stacking, while LEED from a region containing multiple terraces has an averaged 6-fold symmetry \cite{elgabaly}.  Similar measurements can be performed with greater spatial resolution by using dark-field low-energy electron microscopy (LEEM) imaging.  
Dark-field LEEM images are real-space images derived from higher order diffraction spots.  This differs from bright field LEEM, where the images are obtained from the specularly reflected beam, the (0,0) diffraction spot. 
Thus, dark field LEEM can be viewed as a tool comparable to LEED, where the dark field LEEM image is a map of the intensity of a single LEED spot as a function of sample position.  Combining several such images obtained on inequivalent diffraction spots, one can determine direct evidence of asymmetries in the LEED diffraction peaks as a function of position in the LEEM image.

In this Rapid Communication we characterize the crystallographic structure of graphene/SiC films and the spatial distribution of stacking domains by high resolution dark-field LEEM imaging. 
We find that the 6-fold symmetry is broken for the 1x1 SiC LEED spots, and for the 1x1 graphite LEED spots of multilayer ($\geq$2 graphene layers in addition to the buffer layer) graphene.  On the contrary, the apparent 6-fold symmetry of the graphite LEED spots is preserved in the buffer layer and single-layer graphene, showing that the stacking between these two layers differs from that of bilayer graphene.  Interestingly, we also observe that the $6\times6$ satellite spots possess 3-fold symmetry for every measured film thickness.  These measurements of diffraction symmetry help us to understand the properties of epitaxial graphene films.

Atomically-thin graphene samples have been epitaxially grown on the Si-terminated face of 6H-SiC, as previously reported \cite{Rollings}.
The first graphitic layer that forms is a carbon-rich ``buffer layer'' \cite{Mattausch}, which has the same $\sigma$ bands as graphene but the conical dispersion of the graphene $\pi$ bands is absent \cite{Varchonsimulation, Seyller}.
The second graphitic layer is single-layer graphene, with a band gap at the Dirac point due to the graphene substrate interaction \cite{ZhouGap}.
Low energy electron microscopy (LEEM) measurements were performed at the National Center for Electron Microscopy at the Lawrence Berkeley National Laboratory to monitor and characterize the in situ growth \cite{OhtaLEEM,Hibino,Siegel}.  Dark-field LEEM and low energy electron diffraction (LEED) were performed at Sandia National Laboratory to study the crystallographic structure of these samples.



\begin{figure}
\includegraphics[width=8cm] {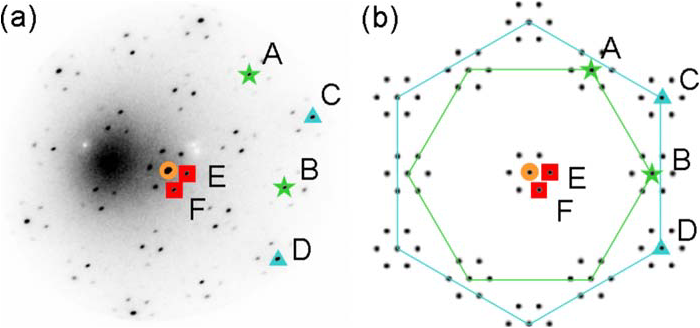}
\caption{(a) A LEED image (43eV) of an epitaxially-grown graphene sample.  The large bright region to the left of center is due to secondary electrons.  (b) A cartoon of the LEED image in (a) that shows the relevant sets of diffraction spots more clearly.  In both panels, two SiC spots are marked with green stars, two graphite spots are marked with blue triangles, two $6\times6$ spots are marked with red squares, and the (0,0) spot is marked with an orange circle.}
\end{figure}

Figure 1a shows a typical LEED pattern \cite{Forbeaux} of a graphene sample with single layer graphene and buffer layer exposed.  Due to the mismatch between the graphene and the SiC lattices, there is a ($6\sqrt(3)\times6\sqrt(3)$)R30$^\circ$ unit cell.  This unit cell appears in the diffraction pattern as bright 6$\times$6 satellite spots around the specular beam and the first order diffraction spots that correspond to the graphite and SiC lattice periodicities.  
These satellite peaks, like the SiC spots, become weaker with increasing graphene film thickness.  

\begin{figure} \includegraphics[width=8.5cm]{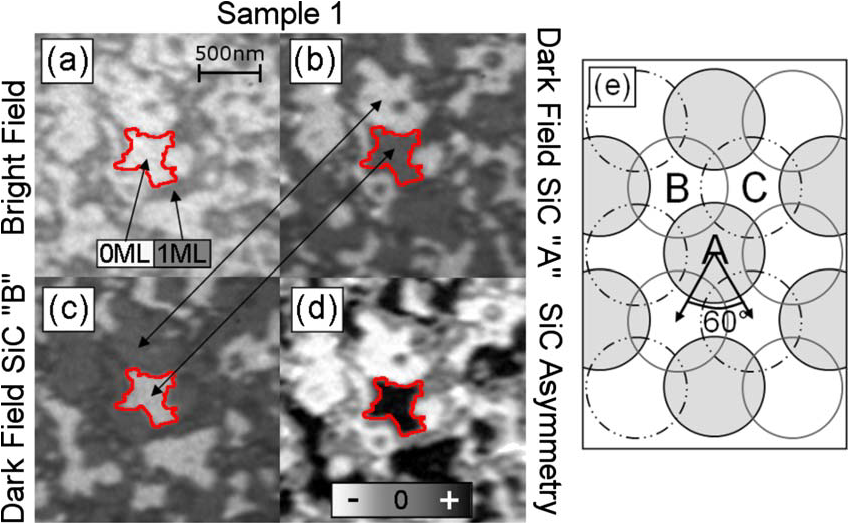}
\caption{2$\mu$m $\times$ 2$\mu$m LEEM images, taken at the same place on sample 1.  (a) A bright field image (3.7eV) where the buffer layer is light grey and single layer graphene is medium-grey.  (b) Dark field image (53.0eV) for the SiC LEED spot labelled ``A'' in figure 1.  (c) Dark field image (also 53.0eV) for the SiC LEED spot labelled ``B'' in figure 1.  (d) Dark field contrast between the two 53.0eV SiC LEED spots shown; panel (d) is a subtraction of panel (c) from panel (b).  In panel (d), positive and negative contrast are given by the black and white regions of the image, and regions of zero contrast are grey.  (e) Cartoon illustrating two ways to stack one layer above another in a closest-packed configuration; AB-stacking differs from AC-stacking by a 60-degree rotation.}\end{figure}

\begin{figure} \includegraphics[width=8.5cm]{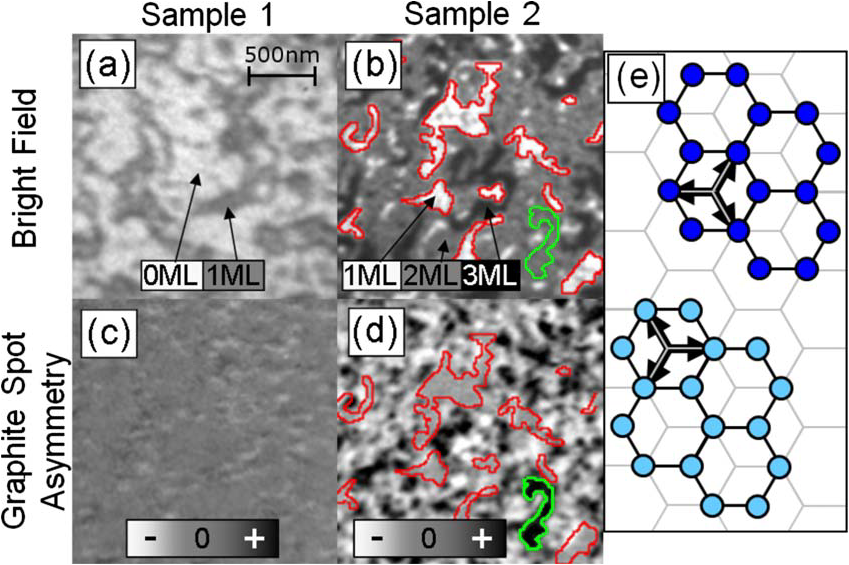}
\caption{2$\mu$m $\times$ 2$\mu$m LEEM images.  (a,b) Bright field images of two different samples.  The grey scale of (a) is different from (b); (a) was taken at 3.7eV, where the buffer layer is light grey, and single layer graphene is medium-grey.  (b) was taken at 5.4eV, where single-layer graphene is light grey or white (rippled-looking), bilayer graphene is medium-grey, and 3-layer graphene is black.  The buffer layer would be a bright white.  (c,d) Dark field contrast images from two graphite LEED spots (labelled ``C'' and ``D'' in figure 1), taken at 49.2eV and 55.7eV, respectively.  Positive and negative contrast are given by the black and white regions of the image, and regions of zero contrast are grey.  Outlines (red for monolayer graphene, green for bilayer) are drawn to help the comparison. (e) Cartoon that shows the two types of stacking domains in panel (d).}\end{figure}


Figure 2 shows dark field LEEM images from SiC LEED spots.  The bright field image, obtained from the (0,0) diffraction spot (central orange circle in figure 1), is shown in figure 2a and provides an accurate determination of the sample thickness by monitoring the intensity contrast as a function of electron energy (Ref. \cite{OhtaLEEM,Hibino,Siegel}).  The dark field images from two SiC LEED spots (spots ``A'' and ``B'' in Fig. 1) are shown in figure 2b and 2c. 
The direct comparison between the two dark field images clearly shows that there are regions of the sample where the intensity is $\it{reversed}$ from one LEED spot to the next (compare e.g., the regions outlined in panels b-c).  
Such intensity change is more obvious when plotting the intensity difference (Fig. 2d) as an ``asymmetry contrast image'', obtained by subtracting panel (b) from panel (c).  This intensity contrast is a direct measure of the asymmetry between two LEED spots.  Regions that look black or white in panel (d) represent the areas of larger asymmetry, while grey in panel (d) corresponds to the regions of almost zero asymmetry.
This asymmetry reflects a 3-fold, rather than 6-fold diffraction symmetry, likely due to the three-fold symmetry of the SiC surface stacking.  A stepped and terraced 6H SiC(0001) surface has, on average, 6-fold crystallographic symmetry, but a single atomic terrace on the SiC surface has 3-fold symmetry. The symmetry relationship between adjacent terraces separated by a step can be understood by considering the stacking sequence of layers perpendicular to the surface in 6H SiC --- ABCACB. Consider two terraces separated by a 3-layer step. If one terrace has AB termination, the other has AC termination.  Since the AB-stacking is 60-degrees rotated from AC-stacking, the diffraction patterns from the two terraces are also 60-degrees rotated.  
This is why anywhere in panel (a) that has buffer layer is either black or white in panel (d).  In contrast, the intensity of diffraction from the single-layer graphene into the SiC LEED spots is significantly lower, which limits our ability to determine the symmetry; this is likely why monolayer graphene in panel (d) is grey.  A summary of these results is given in table 1.


Figure 3 presents the diffraction asymmetry from the graphite spots for monolayer and multilayer graphene from two samples.  The data show that the apparent 6-fold symmetry is preserved only for the buffer layer and monolayer graphene, while it is clearly broken for multilayer graphene ($\geq$2 graphene layers plus buffer layer).  The local film thickness is determined from the bright field images of panels (a) and (b).  Panels (c) and (d) show the asymmetry contrast image obtained from the graphene diffraction spots (labelled ``C'' and ``D'' in figure 1).  In the case of buffer layer (panel c) and monolayer graphene (panel c and d) we do not observe contrast in the asymmetry image for any electron energy studied, despite the presence of a strong signal.  We note that, in comparison to panel (d), there are some regions of faint contrast in panel (c), but these artifacts are at the boundaries of domains and result from imperfections in the image subtraction process.
This lack of contrast indicates that monolayer graphene mantains a 6-fold diffraction pattern.  The lack of 3-fold symmetry is surprising since one might expect the monolayer graphene to be Bernal stacked above the buffer layer.  Furthermore, theoretical calculations predict 3-fold symmetry \cite{Kim}.  In the simplest interpretation, the apparent six-fold crystal symmetry we observe rules out uniform AB stacking (as in the case of bilayer graphene), or even an average A-B asymmetry.  There are, however, other possibilities. First, 3-fold (AB) stacking of the graphene sheets may occur but with a domain size (lateral extent) smaller than the resolution of the LEEM (approximately 10nm).  The average stacking (e.g., AB plus AC) would then have 6-fold symmetry.  Second, graphene may not be Bernal stacked above the buffer layer.   For example, it would be 6-fold symmetric if the carbon atoms of the first graphene monolayer were positioned directly above the carbon atoms of the buffer layer (AA stacking).

\begin{figure} \includegraphics[width=7cm]{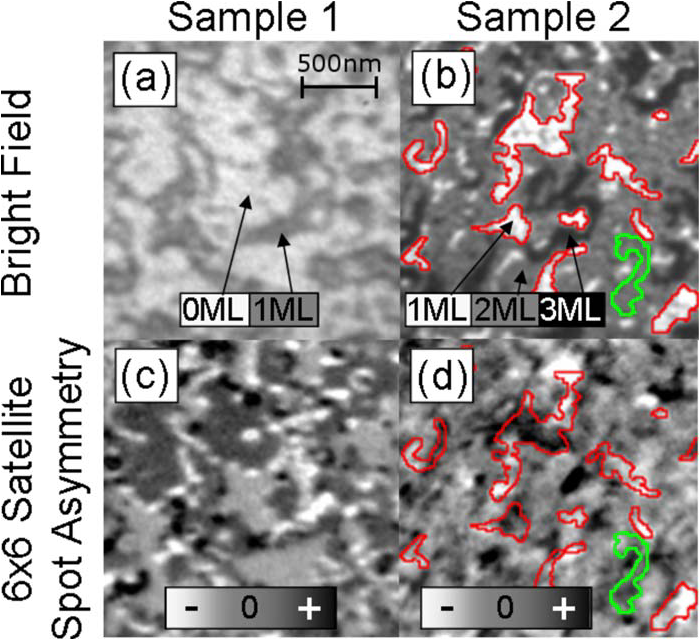}
\caption{2$\mu$m $\times$ 2$\mu$m LEEM images.  (a,b) Bright field images of two samples (the same images as panels (a,b) of figure 3).  (a) was taken at 3.7eV, where the buffer layer is light grey, and single layer graphene is medium-grey.  In (b), taken at 5.4eV, single-layer graphene is light grey or white (rippled-looking), bilayer graphene is medium-grey, and 3-layer graphene is black.  (c, d) Dark field contrast of two $6\times6$ LEED spots, taken at 39.6eV and 14.5eV, respectively.  Positive and negative contrast are given by the black and white regions of the image, and regions of zero contrast are grey.  Outlines (red for monolayer graphene, green for bilayer) are drawn to help the comparison.}\end{figure}

On the contrary, within each 2- and 3-layer region, the asymmetry contrast image reveals regions of black and white contrast suggesting the presence of 3-fold symmetry in these regions. 
These black and white regions correspond to stacking domains with sizes on the order of 100nm for our samples, and are smaller than the domains of uniform thickness seen in the bright-field images (figure 3a and b).
The cartoon in panel (e) of figure 3 shows two ways to stack a second layer of graphene (light and dark blue) on top of a first layer (grey).  The two resulting domains differ from each other by a 60 degree rotation with respect to the graphene layer beneath them (shown by the orientation of the arrows in the cartoon). 
Where the stacking domains impinge, a linear defect (domain boundary) occurs. Since the boundary between a black and white region in panel (d) corresponds to a disruption in one or more of the graphene planes, these domains likely play an important role in the transport properties of bilayer and multilayer graphene films.  Thus, the dark-field imaging reveals that stacking domains and their associated domain boundaries occur within regions of otherwise uniformly thick graphene. These defects are in addition to those defects that result from changes in graphene thickness, and illustrate the complexity that can occur in graphene synthesized from thermally decomposing SiC.


Figure 4 shows the diffraction asymmetry from the $6\times6$ satellite spots, which also possess 3-fold symmetry (a summary of diffraction symmetries is given in table 1).  In the asymmetry contrast image of panel (c), the buffer layer has shades of light and dark grey, and the single layer graphene has regions of black and white. Comparing panel 4(c) to figure 2(d) (the same region of the sample), the regions of asymmetry in the buffer layer are the same for both the satellite spots and the SiC spots.  This observation suggests that the asymmetry in the satellite spots arises in part from the 3-fold symmetry of the SiC substrate.  In fact, LEED from one-layer graphene on a single Ru(0001) terrace \cite{McCarty} is 3-fold symmetric, as are some patterns from graphene on SiC \cite{Johansson}, likely because the substrate terraces were of similar size to the micro-LEED beamspot.

It is also possible that this asymmetry is a reflection of a more complex periodicity.  For example, if the rippling of the thin graphene overlayers caused by the ($6\sqrt(3)\times6\sqrt(3)$)R30$^\circ$ unit cell has a 3-fold nature, the black and white regions in panels (c) and (d) could correspond to 3-fold symmetry domains in the ripples of the overlayer films.  Such rippling has been indicated theoretically \cite{Kim} and experimentally \cite{Varchon}, and is believed to be responsible for variations in the size of the monolayer graphene bandgap \cite{Vitali}.

\begin{table}[h]
\caption{Summary of Diffraction Symmetries} 
\centering     
\begin{tabular}{c cccc}  
\hline\hline                        
Type of Diffraction Spot   &\multicolumn{4}{c}{Graphene Thickness} \\ [0.5ex]
\hline
 & Buffer Layer & 1-Layer & 2-Layer & 3-Layer\\
\hline                
SiC   & 3-Fold & -- & -- & -- \\  
Graphene & 6-Fold & 6-Fold &  3-Fold & 3-Fold \\
Satellite     & 3-Fold & 3-Fold &  3-Fold & 3-Fold \\[1ex] 
\hline                          
\end{tabular}
\label{tab:hresult}
\end{table}

In conclusion, we have studied the crystallographic structure of epitaxial graphene.  We find that there is a fundamental difference in the stacking of the buffer layer, monolayer and bilayer graphene on top of the SiC substrate. We show that the structure of the buffer layer is more complex than previously expected, and that dark field LEEM provides a new method to directly characterize the domain sizes of bilayer graphene and to extract information about the structure of the SiC interface beneath few-monolayer graphene films.  We anticipate that the ability to image these crystal symmetries will continue to enhance our understanding of the properties of thin graphene films on various substrates in the future.

We thank D.-H. Lee for useful discussions. LEEM measurements
and sample growth were supported by the Division
of Materials Sciences and Engineering of the U.S
Department of Energy under Contracts No. DEAC03-
76SF00098 (LBL) and DE-AC04-94AL85000 (SNL). Sample growth
was also supported by the MRSEC project through the National Science Foundation.

Correspondence and requests for materials should be addressed to Alanzara@lbl.gov.

\newpage
$
\\
\\
\\
\\
\\
\\
\\
$
\part*{Addendum for the arXiv version}
During the course of our study, we found interesting spiral patterns on hydrogen-etched substrates.  For the arXiv version we have added them below, since they contribute towards a general understanding of the structure of SiC(0001), as well as our experimental technique and analysis method.  Note that this is not a part of our original publication.

\newpage

During the course of this study, we found interesting spirals caused by screw dislocations on hydrogen-etched 6H-SiC(0001).  This sample was grown by the Walt deHeer group at Georgia Tech and did not contribute to the original publication, but the data were measured simultaneously.  Such defects have been seen and studied in the past on similar samples; some nice AFM images can be seen, for example, in Ref \cite{niceAFM}.  The Burgers vectors point in or out of the page.

\begin{figure*} \includegraphics[width=16cm]{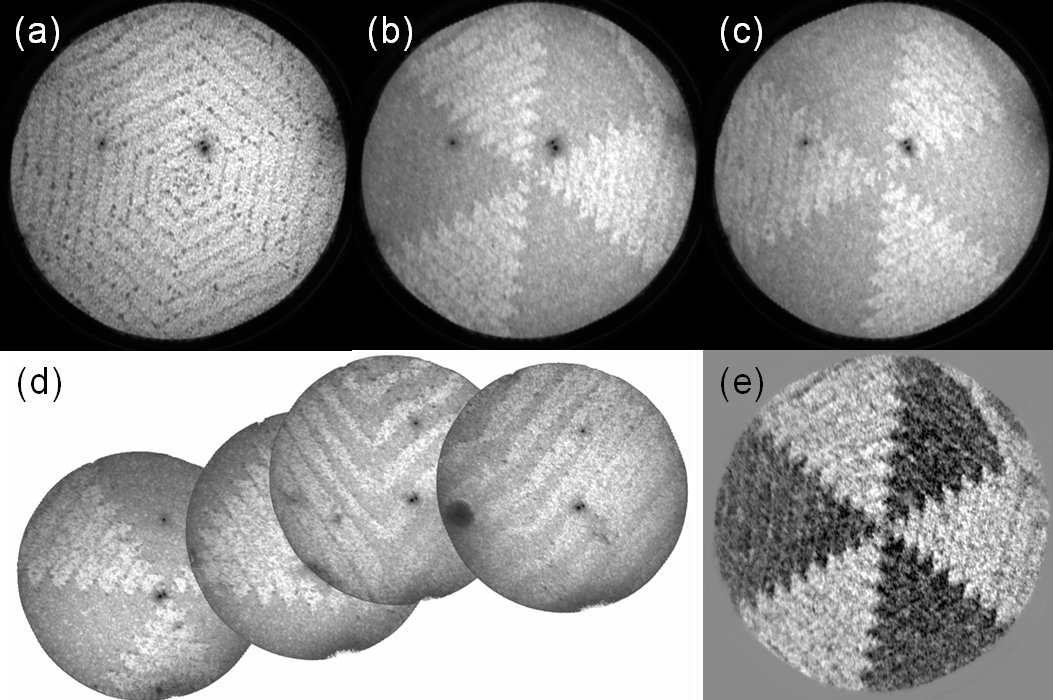}
\caption{14$\mu$m LEEM images (arXiv version only).}\end{figure*}

Panel (a) shows a bright field LEEM image with a 14 $\mu$m field of view, taken at a screw dislocation.  This sample was partially graphitized (darker grey), but most of this image is buffer layer (lighter grey).  The two dark spots near the center of the image are artifacts.

Panels (b) and (c) show two dark field LEEM images, taken on the same spot of the sample but imaged with different 6$\times$6 spots (corresponding to ``E'' and ``F'' of our original publication).  Since the SiC substrate is 3-fold symmetric due to its ABCACB layer order, the crystal symmetry produces 3-fold LEED patterns and dark field images with two types of contrast (ABC vs ACB stacking).

Panel (d) is similar to (b) or (c), but has several 14 $\mu$m views stitched together.  Near the center of the spiral (along the thick red line), adjacent steps are the same shade of light gray; whereas further away from the center (along the thinner blue line), adjacent steps are alternating shades of light and dark gray.  This shows that step bunching contributes to the geometrical pattern in the vicinity of the defect.

Panel (e) shows the asymmetry contrast between two 6$\times$6 diffraction spots.  A similar image could have been made by taking the asymmetry contrast from two SiC spots.  The image in panel (e) is analogous to figures 4c and 4d of the original publication.

These images give another demonstration that the examination of crystal symmetries can contribute to the understanding of atomic layer stackings at the surface of a crystal.

\end{document}